\documentclass[submission, PhysCodeb]{SciPost_better_arXiv}

\hypersetup{
    colorlinks,
    linkcolor={red!50!black},
    citecolor={blue!50!black},
    urlcolor={blue!50!black}
}

\usepackage{pdfcomment}
\usepackage{sansmath} % sans serif math

\DeclareSymbolFont{usualmathcal}{OMS}{cmsy}{m}{n}
\DeclareSymbolFontAlphabet{\mathcal}{usualmathcal}

\usepackage{booktabs,tabularx}
\usepackage{xspace}
\usepackage{subcaption}

\usepackage[utf8]{inputenc} 
\usepackage[compat=1.0.0]{tikz-feynman}
\usepackage{scalerel}
\usepackage{layouts}
\usepackage{soul}

\usepackage{tikz-layers}
\usepackage{lineno}
\usepackage{printlen}
\usepackage{wasysym}
\usepackage{grffile}

\newcommand{\as}{\alpha_s}

\newcommand{\MSbar}{\ensuremath{\overline{\text{MS}}}\xspace}

\newcommand{\nc}{N_\text{\textsc{c}}} 
\newcommand{\order}[1]{\mathcal{O}\left(#1\right)}

\newcommand{\logbook}[2]{}
\newcommand{\betaps}{{\beta_\text{\textsc{ps}}}} 
\newcommand{\betaobs}{{\beta_\text{obs}}} 

\newcommand{\panscales}{\textsc{PanScales}\xspace}
\newcommand{\pythia}{\textsc{Pythia8.3}\xspace}
\newcommand{\rivet}{\textsc{Rivet}\xspace}

\newcommand{\ttt}[1]{\texttt{#1}}
\newcommand{\repolink}[2]{\href{https://gitlab.com/panscales/panscales-0.X/-/blob/main/#1#2}{\ttt{#2}}}
\newcommand{\showercodelink}[1]{\href{https://gitlab.com/panscales/panscales-0.X/-/blob/main/shower-code/#1}{\ttt{#1}}}
\newcommand{\analysislink}[1]{\href{https://gitlab.com/panscales/panscales-0.X/-/blob/main/analysis/nll-validation#1}{\ttt{#1}}}
\newcommand{\mainlink}[1]{\href{https://gitlab.com/panscales/panscales-0.X/-/blob/main/#1}{\ttt{#1}}}
\newcommand{\ourpythialink}[1]{\href{https://gitlab.com/panscales/panscales-0.X/-/blob/main/pythia-interface/#1}{\ttt{#1}}}
\newcommand{\email}[1]{\href{mailto:#1}{#1}}

\definecolor{darkgreen}{rgb}{0,0.4,0}
\definecolor{grey}{rgb}{0.5,0.5,0.5}
\definecolor{orange}{rgb}{0.9,0.5,0.0}
\definecolor{lightblue}{rgb}{0.0,0.5,1.0}
\definecolor{darkspringgreen}{rgb}{0.09, 0.45, 0.27}

\begin{document}
\begin{flushright}
CERN-TH-2023-238, Nikhef 2023-026, OUTP-23-16P
\end{flushright}
\begin{center}{\Large \textbf{
      Introduction to the PanScales framework, version 0.1
\\ }}\end{center}

\begin{center}
  Melissa van Beekveld\textsuperscript{1},
  Mrinal Dasgupta\textsuperscript{2},
  Basem Kamal El-Menoufi\textsuperscript{2,3},
  Silvia Ferrario Ravasio\textsuperscript{4},
  Keith Hamilton\textsuperscript{5},
  Jack Helliwell\textsuperscript{6},
  Alexander Karlberg\textsuperscript{4},
  Rok Medves\textsuperscript{6},
  Pier Francesco Monni\textsuperscript{4},
  Gavin P. Salam\textsuperscript{6,7},
  Ludovic Scyboz\textsuperscript{3,6},
  Alba Soto-Ontoso\textsuperscript{4},
  Gregory Soyez\textsuperscript{8},
  Rob Verheyen\textsuperscript{5}
\end{center}

\begin{center}
  {\small
{\bf 1} Nikhef, Theory Group, Science Park 105, 1098 XG, Amsterdam, The Netherlands \\
{\bf 2} Department of Physics \& Astronomy, University of Manchester, Manchester M13 9PL, United Kingdom \\
{\bf 3} School of Physics and Astronomy, Monash University, Wellington Rd, Clayton VIC-3800, Australia\\
{\bf 4} CERN, Theoretical Physics Department, CH-1211 Geneva 23, Switzerland \\
{\bf 5} Department of Physics and Astronomy, University College London, London, WC1E 6BT, UK \\
{\bf 6} Rudolf Peierls Centre for Theoretical Physics, Clarendon Laboratory, Parks Road, University of Oxford, Oxford OX1 3PU, UK \\
{\bf 7} All Souls College, Oxford OX1 4AL, UK \\
{\bf 8} Universit\'e Paris-Saclay, CNRS, CEA, Institut de physique th\'eorique, 91191, Gif-sur-Yvette, France
\\
  {\small \sf \email{mbeekvel@nikhef.nl},
    \email{mrinal.dasgupta@manchester.ac.uk},
    \email{basem.el-menoufi@monash.edu},
    \email{silvia.ferrario.ravasio@cern.ch},
    \email{keith.hamilton@ucl.ac.uk},
    \email{jack.helliwell@physics.ox.ac.uk},
    \email{alexander.karlberg@cern.ch},
    \email{pier.monni@cern.ch},
    \email{gavin.salam@physics.ox.ac.uk},
    \email{ludovic.scyboz@monash.edu},
    \email{alba.soto.ontoso@cern.ch},
    \email{gregory.soyez@ipht.fr}
  }}
\end{center}

% \begin{center}
% \today
% \end{center}

% For convenience during refereeing (optional),
% you can turn on line numbers by uncommenting the next line:
%\linenumbers
% You should run LaTeX twice in order for the line numbers to appear.

\section*{Abstract}
{\bf
In this article, we document version 0.1 of the \panscales code
for parton shower simulations.
With the help of a few examples, we discuss basic usage of the code,
including tests of logarithmic
accuracy of parton showers.
We expose some of the numerical techniques underlying the logarithmic
tests and include a description of how users can implement their own
showers within the framework.
Some of the simpler logarithmic tests can be performed in a few minutes on a modern laptop.
As an early step towards phenomenology, we also outline some aspects
of a preliminary interface to \pythia, for access to its hard matrix
elements and its hadronisation modules.

\begin{center}
  The code is available from
  \url{https://gitlab.com/panscales/panscales-0.X}
\end{center}

}
\newpage

% TODO: include a table of contents (optional)
% Guideline: if your paper is longer that 6 pages, include a TOC
% To remove the TOC, simply cut the following block
\vspace{10pt}
\noindent\rule{\textwidth}{1pt}
\tableofcontents\thispagestyle{fancy}
\noindent\rule{\textwidth}{1pt}
\vspace{10pt}
\section{Introduction}
\label{sec:intro}

Parton showers lie at the core of the majority of experimental and
phenomenological studies in collider physics. At the LHC, they connect
the electroweak and TeV momentum scales of hard-scattering processes, where the
relevant degrees of freedom are perturbative quarks and gluons, with
the non-perturbative physics of hadrons at scales of a few hundred
MeV.
As such, parton showers account for physics across several orders of
magnitude in momentum scales.
In QCD, large logarithms typically appear in the presence of large
momentum scale hierarchies, which have to be resummed to all orders in
the strong coupling to obtain physically sensible results. 
One of the frontiers of the development of parton showers is to
understand, demonstrate and improve their logarithmic accuracy, with
analytic resummations providing crucial inputs, as well as reference
results for comparison.

This paper documents the first public release of a new parton
showering code, \panscales, version 0.1.
It has been developed as part of a series of
articles~\cite{Dasgupta:2020fwr,Hamilton:2020rcu,Karlberg:2021kwr,Hamilton:2021dyz,vanBeekveld:2022zhl,vanBeekveld:2022ukn,Hamilton:2023dwb,vanBeekveld:2023lfu,FerrarioRavasio:2023kyg}
investigating how to design parton shower algorithms that provide
controlled and verifiable logarithmic accuracy, together with parallel
analytical work on approaches to resummation at higher logarithmic
accuracy and their connection with parton
showers~\cite{Dasgupta:2018nvj,Dasgupta:2021hbh,Banfi:2021owj,Banfi:2021xzn,Dasgupta:2022fim,Medves:2022ccw,Medves:2022uii,vanBeekveld:2023lsa}.
Several other groups have also recently been working on the question
of logarithmic accuracy in showers, see e.g.\
Refs.~\cite{Hoche:2017kst,Bewick:2019rbu,Bewick:2021nhc,Forshaw:2020wrq,Nagy:2020rmk,Nagy:2020dvz,Herren:2022jej,Assi:2023rbu}.

This \panscales release includes
two main NLL-accurate parton showers, PanGlobal and PanLocal.
They have had their next-to-leading-logarithmic
(NLL) accuracy tested for $e^+e^-$~\cite{Dasgupta:2020fwr} and
colour-singlet production in
$pp$~\cite{vanBeekveld:2022zhl,vanBeekveld:2022ukn} collisions, as
well as Deep Inelastic Scattering (DIS) and Vector Boson Fusion (VBF)
processes~\cite{vanBeekveld:2023lfu}.
They include state-of-the-art handling of subleading colour
corrections, which for many processes and observables allows for full colour accuracy
at LL (and often beyond)~\cite{Hamilton:2020rcu,vanBeekveld:2022zhl}. 
They also include the treatment of both collinear and soft spin
correlations~\cite{Karlberg:2021kwr,Hamilton:2021dyz,vanBeekveld:2022zhl},
and first steps towards matching~\cite{Hamilton:2023dwb}, as well as
elements towards NNLL accuracy~\cite{FerrarioRavasio:2023kyg} (the
latter two just for $e^+e^-$ collisions).
Finally, the codebase contains early versions of features that are yet
to be discussed in physics research papers, in particular an interface
with the \pythia event generator~\cite{Bierlich:2022pfr}, which can be
used to provide hard-process generation and hadronisation.
Despite the inclusion of the interface to \pythia, the code is not yet
at a stage of maturity that is suitable for extensive comparisons to
experimental data.
This is notably because of the absence of finite quark-mass effects,
the need for further work on matching with higher (fixed) order
effects, as well as tuning of the non-perturbative parameters of the
shower and of any hadronisation model with which it is used.

This manuscript is structured as follows.
Section~\ref{sec:basic-usage} focuses on basic usage of the code,
illustrating: the build procedure (section~\ref{sec:get-and-build});
stand-alone event generation (section~\ref{sec:standalone});
the use of the code for carrying out basic logarithmic tests of parton
showers (section~\ref{sec:running-log-tests});
usage with a preliminary interface to \pythia
(section~\ref{sec:pythia-usage});
and details for carrying out validation of the code and building it
with higher numerical precision
(section~\ref{sec:advanced-builds-plus-validation}).
Section~\ref{sec:cons-log-tests} illustrates some of the techniques
that underlie the logarithmic tests, discussing both
double-logarithmic global event shape observables
(section~\ref{sec:log-tests-global}) and single-logarithmic 
non-global observables (section~\ref{sec:non-global-log-tests}).
Section~\ref{sec:new-shower} gives a brief discussion of how to use
the PanScales framework to implement a new shower, which provides a
relatively straightforward way to gain access to the colour,
spin-handling and logarithmic-accuracy testing facilities.
We close in section~\ref{sec:outlook} with an outlook.

%======================================================================
\section{Basic usage}
\label{sec:basic-usage}

The \panscales code requires a C++14 compiler and a Fortran~95 compiler,
the GSL library, CMake ($\ge 3.7$) and, for
some scripts, \texttt{Python} ($\ge 3.6$) with \ttt{matplotlib} installed.
Some features (higher-precision builds) require the
MPFR~\cite{MPFR} and QD~\cite{hida2000quad} libraries (see
section~\ref{sec:advanced-builds-plus-validation} for details).
It includes several third party codes, notably
\ttt{fjcore}~\cite{Cacciari:2011ma} for jet finding and
\ttt{hoppet}~\cite{Salam:2008qg} for PDF handling and the
\href{https://github.com/catchorg/Catch2}{Catch2} library for
unit-testing (see the \mainlink{3rdPartyCode.md} file for further
details).
The code can also be linked to LHAPDF~\cite{Buckley:2014ana}
and to \pythia.
The \panscales code is released under the GNU GPL v3 license.

%----------------------------------------------------------------------
\subsection{Downloading and building the code}
\label{sec:get-and-build}

\panscales can be obtained from the git repository
\begin{verbatim}
  git clone --recursive https://gitlab.com/panscales/panscales-0.X 
\end{verbatim}
The main code is in the \mainlink{shower-code/} subdirectory. 
To build the code and examples in double-precision, do the following
\begin{verbatim}
  cd panscales-0.X/shower-code
  ../scripts/build.py -j
\end{verbatim}
The \mainlink{scripts/build.py} script uses CMake to organise the build,
which by default is placed in the \ttt{build-double/} subdirectory.
Advanced use of CMake is described in the \showercodelink{BUILD.md}
file.

% ----------------------------------------------------------------------
\subsection{Standalone event generation}
\label{sec:standalone}

To run showering for the $e^+e^- \to q\bar q$ process and analyse the events, do
\begin{verbatim}
  build-double/example-ee -shower panglobal -beta 0 -process ee2qq \
        -physical-coupling -rts 91.1876 -nev 100000 \ 
        -out example-results/example-ee.dat
\end{verbatim}
This will take about $5{-}10$ seconds, and produce an output file with
histograms for a range of event shapes.
Runs are generally configured with command-line options, for example
with the first line indicating the use of the PanGlobal shower.
The shower ordering variable is $v = k_t e^{-\betaps |\eta|}$, with $k_t$ and
$\eta = - \ln \tan \theta/2$ respectively the transverse momentum and
pseudorapidity of the emission with respect to its parent.
The choice $\betaps=0$ (set with the \ttt{-beta 0} command-line
argument) corresponds to transverse-momentum ordering.

The available command-line options can be explored with the
\ttt{-h} flag, and command-line options can also be placed in a
card-file and read with the \ttt{-argfile card-file.txt} option.
For convenience, the main options are also listed in an
\showercodelink{OPTIONS.md} file (any executable can be made to
generate such a file by adding \ttt{-markdown-help} to the command
line).
More details on the \panscales interface can be found by examining the
\showercodelink{example-ee.cc} file.
Much of the code also has \ttt{doxygen} documentation, which can be
obtained by running \ttt{doxygen} from the \mainlink{shower-code/}
directory.

Note that in standalone mode, as given above, currently all events
have unit weight, i.e.\ in order to recover a physical cross-section
from the above run the histograms should be multiplied by the
appropriate cross section for the hard process.
Furthermore, in most cases the Born event is a fixed configuration
rather than being sampled over.

A final comment is that the event record (in the class
\ttt{panscales::Event}) holds only the particles as they appear
after all showering, rather than containing all intermediate steps as
in some other codes.
The reason for this is that for showers with global recoil, storing
all intermediate steps would take memory of order $n^2$ for an
$n$-particle event.
By default the code prints the first event, but this can be changed
to print $N$ events with the \ttt{-max-print $N$} option.

Similar examples for $pp \to Z$ and DIS are given respectively in the
\showercodelink{example-pp.cc} and \showercodelink{example-dis.cc},
with further explanations in an \showercodelink{EXAMPLES.md} file.
Note that the above examples do not by default integrate over the
hard-process kinematics.
That functionality is instead available through the interface with
\pythia (cf.\ section~\ref{sec:pythia-usage}).

%----------------------------------------------------------------------
\subsection{Logarithmic tests}
\label{sec:running-log-tests}

Within the \mainlink{shower-code/} directory, the
\showercodelink{example-global-nll-ee.py} script, and associated
\showercodelink{example-global-nll-ee.cc} program, illustrate how to carry out a
basic test of the NLL accuracy of a shower for global event-shape
observables.
It can be used as 
\begin{verbatim}
  ./example-global-nll-ee.py --njobs NJOBS --shower panglobal
\end{verbatim}
and will test the NLL accuracy of the \(e^{+}e^{-}\) PanGlobal shower with
$\betaps=0$ (i.e.\ $k_t$ ordered), for event shapes like the Cambridge-algorithm
\(y_{23}\)~\cite{Dokshitzer:1997in} and Lund
observables~\cite{Dasgupta:2020fwr}.
The Lund observables measure either the sum or the maximum of the
$k_{ti} e^{-\betaobs |\eta_i|}$ for primary Lund-plane declusterings
$i$~\cite{Dreyer:2018nbf}.
The NLL test works with $\betaobs = 0$, i.e.\ examining just the
relative transverse momenta ($k_t$) of the declusterings.

\begin{figure}
  \centering
  \includegraphics[width=0.32\textwidth]{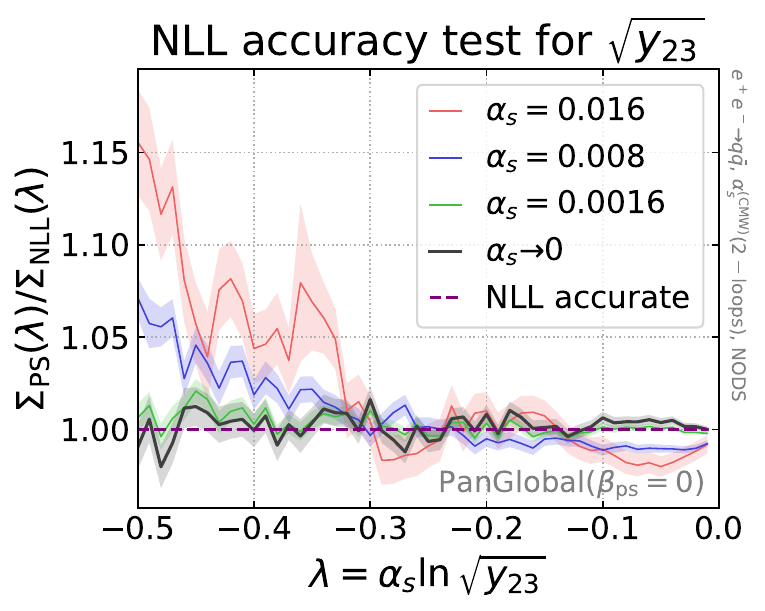}
  \includegraphics[width=0.32\textwidth]{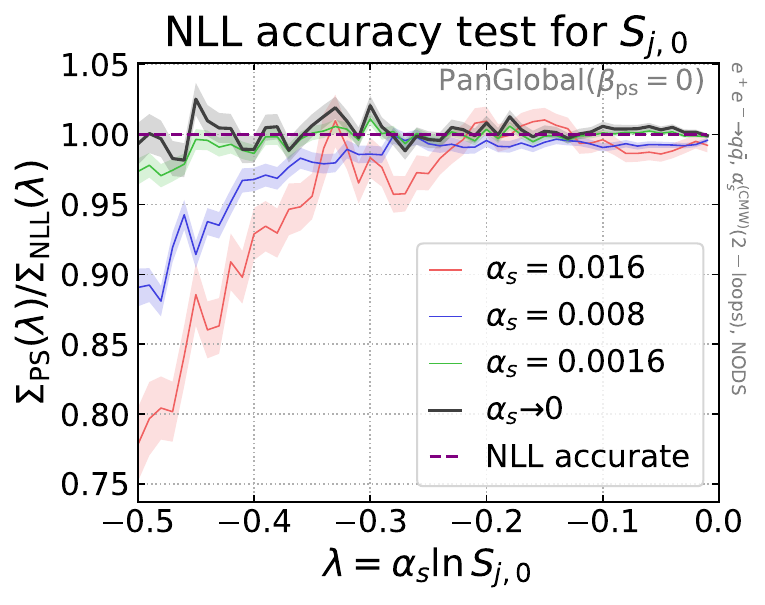}
  \includegraphics[width=0.32\textwidth]{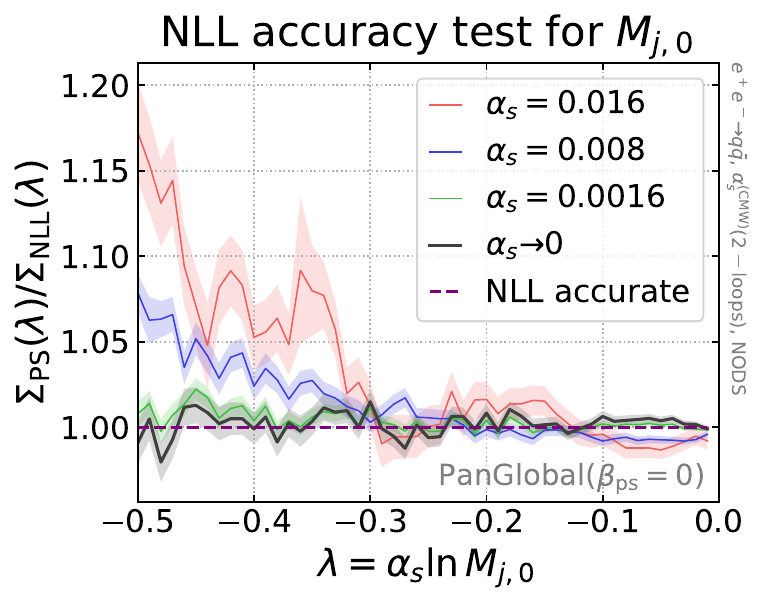}
  \caption{Example results from tests of NLL accuracy of the PanGlobal
    ($\betaps=0$) shower for three global event-shape
    observables,
    the Cambridge algorithm $y_{23}$ and the Lund $\betaobs=0$ sum and
    maximum observables,
    corresponding respectively to the three panels. 
    Each panel
    shows the
    ratio of the shower cumulative-distribution result to the NLL
    calculation.
    Each coloured line corresponds to a specific value of $\as$, while
    the black line gives the extrapolation to $\as=0$.
    The results are shown as a function of the maximum allowed value
    of $\lambda = \as \ln O$,
    where $O$ is the observable.
  }
  \label{fig:nll-global}
\end{figure}

The \ttt{-\/-njobs} flag takes an integer which should be equal to the
number of cores that one wishes to run on. 
Depending on the machine, the script takes around 15 to 50 CPU minutes,
i.e. a few minutes of wall-time on a modern multi-core
machine.
On completion, for each observable, it will produce plots such as
those in Fig.~\ref{fig:nll-global}, showing the $\as \to 0$ limit of the ratio 
$\Sigma_\text{shower}(\lambda)/\Sigma_\text{NLL}(\lambda)$, where $\Sigma(\lambda)$ is
the cross section for $\as \ln O$ to be smaller than $\lambda$, with
$O$ being the value of the observable.
The plot is given as a function of $\lambda$,
as is standard for NLL
tests~\cite{Dasgupta:2020fwr}: for an NLL-correct
shower, the $\as \to 0$ limit of the ratio will be equal to $1$.
The user can examine the results for a non-NLL shower by replacing
\ttt{panglobal} on the command line with \ttt{dipole-kt}, which
provides a standard $k_t$-ordered dipole
shower~\cite{vanBeekveld:2022zhl}, much like
those~\cite{Sjostrand:2004ef,Schumann:2007mg,Platzer:2009jq} in
standard public tools.

The \mainlink{shower-code} directory also includes an
\showercodelink{example-nonglobal-nll-ee.py} script for testing
non-global logarithms in the context of energy flow into an angular
slice. 
It can be used as follows
\begin{verbatim}
  ./example-nonglobal-nll-ee.py --njobs NJOBS --shower panglobal
\end{verbatim}
The script again generates a plot with the ratio to the NLL
(single-logarithmic) result.

The above command lines serve mainly to illustrate the more
straightforward logarithmic accuracy tests and provide only a subset
of the functionality required for a full set of tests.
For example, the global event-shape tests in the
\showercodelink{example-global-nll-ee.py} script are limited to
$\betaps=\betaobs=0$.
Further discussion of considerations for logarithmic
tests is given in section~\ref{sec:cons-log-tests}.

% ----------------------------------------------------------------------
\subsection{Usage with Pythia8}
\label{sec:pythia-usage}

While we do not yet recommend the \panscales code for phenomenological
production purposes, those wishing to start exploring such
applications may try the interface with the
\pythia\cite{Bierlich:2022pfr} generator code, which we have tested
with version 8.3.10.
This enables the use of  \pythia to generate the hard process,\footnote{So far
  only a limited set of processes is supported in the interface, e.g.\
  because of the setup of subleading-colour information for the hard
  process, which is currently handled manually.} as
well as for hadronisation, while \panscales carries out the parton
showering at the scales in between.
It also enables access to \pythia I/O, e.g.\ for outputting HepMC
files~\cite{Buckley:2019xhk}.

To compile and run the code, enter the
\mainlink{pythia-interface/} directory and run\footnote{Users who wish to
  reuse an existing \pythia installation should see
  \mainlink{pythia-interface/README.md} for
  instructions.
  For users who already have a version of \pythia installed but also
  run the \ttt{get-pythia.sh} script, care should to be taken about conflicts
  between the two versions of \pythia (e.g.\ library paths for dynamic
  linking leading to inconsistent versions).
}
\begin{verbatim}
  ./get-pythia.sh
  ../scripts/build.py --build-lib -j --with-lhapdf
  build-double/main-dy -physical-coupling -lhapdf-set CT14lo \
        -shower panglobal -nev 1e6 -out main-dy.dat
\end{verbatim}
This will simulate Drell-Yan production
and histogram the $Z$ rapidity, mass and transverse momentum.
It does not (yet) include matching to higher-order matrix elements, so
kinematic distributions such as the transverse momentum of the colour singlet $p_{tZ}$ are sensible only in the resummation region,
i.e.\ at low $p_{tZ}$.
This example uses the CT14lo set~\cite{Dulat:2015mca} from
LHAPDF~\cite{Buckley:2014ana},\footnote{The code can also be used
  without LHAPDF, instead using a replacement toy PDF set. See
  \mainlink{pythia-interface/README.md} for details.} and will produce warnings concerning
$x$ regions where the PDF set is badly behaved.
Most
other PDF sets have issues with negative or zero parton
distribution functions, especially at large $x$, that cause the
\panscales code to throw an exception after some number of events.
Ultimately, we intend to make \panscales more tolerant of ill-behaved
PDFs, but at this stage of development we have taken the approach that
it is safer to abort the run than to continue generating events when a
clear problem has arisen, even if only a rare occurrence.

The same directory contains a range of other examples that can be run with  \pythia, and
the header of each example illustrates how to use it.
We also provide an example of interfacing to \rivet~\cite{Bierlich:2019rhm}, in which
case one needs to make sure \rivet is installed and the code
is compiled with it (through \ttt{--cmake-options="-DWITH\_RIVET=on"}). 
This allows for an easy comparison with data, although
we should note that none of the \panscales showers are currently tuned
and do not include the effects of quark masses.

The \panscales-\pythia interface transfers the event into the
\pythia event record after each shower emission.
This is done to have access to the full \pythia functionality in the
future, i.e.\ interleaving
multi-parton interactions (MPI) with showering of the hard process.
Note that at this moment, we do not have the functionality to run
with MPI, but hadronisation can be added through the flag \ttt{-hadron-level}. 

The above examples all use the same \panscales event-loop framework as
in the main \mainlink{shower-code/} directory.
We also distribute examples with a standard \pythia structure.
These are to be found in the \ourpythialink{main-pythia02.cc} and
\ourpythialink{main-pythia06.cc} files, which are based on the
corresponding \ttt{main02.cc} and \ttt{main06.cc} examples from the
\pythia distribution.
Note that only a restricted set of options is supported in this form,
which can be found at the end of
\ourpythialink{PanScalesPythiaModule.cc}. 

%----------------------------------------------------------------------
\subsection{Code validation and more advanced builds}
\label{sec:advanced-builds-plus-validation}

To validate that the code is generating expected results, enter
the \mainlink{shower-code/validation/} directory and run
\begin{verbatim}
  ./validate-showers.py -j
\end{verbatim}
which runs a range of validation tests in parallel across all
available cores.
This takes a total of about $300{-}1000$ CPU seconds on a typical
laptop, carrying out of the order of $100$ separate short runs, each with
different settings, verifying that they give histograms that are
identical to those in a set of reference files.
It is possible to carry out validation runs with larger numbers of
events, but one should be aware that there can be
differences due to varying floating-point behaviours across different
hardware.
Additionally, lower-level unit tests can be found in the
\mainlink{unit-tests/} directory.
The unit tests and a subset of the validation tests can also be run
with the continuous integration script
\mainlink{scripts/CI-build-and-test.py}.

It is sometimes useful to build the main code in different numerical
precisions, e.g.\ for logarithmic tests that probe very disparate
energy scales and angles.
For this, the general build script has an option \ttt{--builds X} which
essentially invokes \ttt{cmake} with a suitable set of
configuration options.
Specifically, one runs
\begin{verbatim}
  ../scripts/build.py --builds X [-j]
\end{verbatim}
where \ttt{X} is a space-separated list that contains one or more of
the following options: \ttt{double}, \ttt{ddreal}, \ttt{qdreal},
\ttt{doubleexp}, \ttt{mpfr4096}.

The \ttt{ddreal} and \ttt{qdreal} options require at least version
2.3.23 of the QD library~\cite{hida2000quad} and have precisions of about
twice and four times that of a double type, with speeds that are about
$10$ and $200{-}300$ times slower.

The \ttt{doubleexp} type was developed specifically for the
\panscales project and has the same relative precision as
\ttt{double}, but a much larger range of exponents (stored in an
additional 64-bit integer), which is useful when exploring finite
values of $\as \ln v$ with very small values of $\as$ and
correspondingly large values of $\ln v$.
It is about $3{-}10$ times slower than \ttt{double}.

The \ttt{mpfr4096} type is based on the MPFR library and has $4096$
bits for the mantissa, i.e.\ about $75$ times higher than double
precision, or equivalently a little over 1000 decimal digits of
precision. 
It is about $1500{-}2500$ times slower than \ttt{double} (using
version 4.2.1 of the library).

For most purposes, the \ttt{double} and \ttt{doubleexp} types are
sufficient, notably when used with methods that track the differences
between directions of dipole ends in addition to the actual momenta of
the corresponding partons~\cite{Hamilton:2020rcu}.
That tracking can be enabled at run time with the \ttt{-use-diffs}
option and has only a modest speed penalty, of the order of
$10\%$.
The higher precision types are, however, important when developing new
showers and testing the correctness of any parts of the code that
carry out the dedicated calculations with direction differences.

% ======================================================================
\section{Further details for logarithmic tests}
\label{sec:cons-log-tests}

In this section we provide some insight into features of the code that
facilitate shower logarithmic accuracy tests, together with a more
detailed discussion of some of the underlying methodology than has
been given in previous work.

Some of the discussion below concerns tests that go beyond the simple
ones illustrated in section~\ref{sec:running-log-tests}.
Code for these more advanced tests is to be found in the directory
\mainlink{analyses/nll-validation/} with usage explanations in the
corresponding \repolink{analyses/nll-validation/}{README.md} file.

%----------------------------------------------------------------------
\subsection{Global event shapes}
\label{sec:log-tests-global}

We start by considering the global event shape tests of
section~\ref{sec:running-log-tests} and specifically examine the
commands that are run by the \showercodelink{example-global-nll-ee.py}
script.
For each of several $\as$ values, that script executes one or more
commands of the following kind

{\footnotesize%
\begin{verbatim}
  build-double/example-global-nll-ee -Q 1.0 -shower panglobal -beta 0.0 \
        -alphas 0.0016 -nloops 2 \
        -lambda-obs-min -0.5 -lnkt-cutoff -327.5 -dynamic-lncutoff -15 \
        -weighted-generation -nev-cache 75000.0 \
        -spin-corr off -use-diffs -nev 750000 -rseq 11 \
        -out example-results/lambda-0.5-alphas0.0016-rseq11.res
\end{verbatim}}
\noindent The second line indicates the value of $\as(Q)$ in the
$\MSbar$ scheme (working
in units of $\sqrt{s}\equiv Q=1$) and that a two-loop running
coupling is to be used.
The next two lines contain two separate critical elements, which we
discuss below in more detail.
Further options are $\ttt{-spin-corr off}$, which turns off spin
correlations (leaving them on adds about $50\%$ to the run time).
The \ttt{-use-diffs} option turns on tracking of direction
differences for higher numerical precision. 
It is not strictly necessary for this example, but the speed cost is
relatively limited, at about $10\%$.
\logbook{958acf794}{
  GPS M2Pro timings:
  - with -use-diffs: 40.67s user 0.77s system 97% cpu 42.349
  - without        : 37.06s user 0.68s system 99% cpu 37.871 total
}%
The \ttt{-rseq 11} argument specifies the random sequence
that is used.

The two critical elements that we now need to discuss in more detail
are (1) a dynamic generation cutoff, which is necessary to prevent
event particle multiplicities from becoming intractable and (2)
weighted event generation, which is necessary in order to explore
regions with large Sudakov suppression.

%......................................................................
\subsubsection{Dynamic generation cutoff}
\label{sec:veto-buffer}

Let us discuss the following part of the command line:
{\footnotesize%
\begin{verbatim}
    -lambda-obs-min -0.5 -lnkt-cutoff -327.5 -dynamic-lncutoff -15 
\end{verbatim}}%
\noindent This indicates that the minimal value of $\lambda = \as \ln v$ is
$-0.5$, which corresponds to $\ln k_t/Q = -312.5$ with $\as = 0.0016$.
The choice \ttt{-lnkt-cutoff -327.5} allows the shower to run somewhat
below the $k_t$ scale associated with the minimal value of $\lambda$.
This is important, because multiple emissions break the immediate
relation between the shower scale and the observable, e.g.\ because of
shower recoil effects, and because the observable sums contributions
from multiple emissions.
These effects are properly accounted for only if the shower is allowed
to evolve sufficiently far below the single-emission scale that is
equivalent to the smallest value of the observable that is probed.

If we were to straightforwardly run with the very small cutoff
indicated above, then the average event particle multiplicity, $n$,
would be prohibitively large.
Specifically, it scales as
$\ln n = |\lambda_c| \sqrt{2C_A/(\pi\as)} +
\order{1}$~\cite{Catani:1991pm}, where $\lambda_c = \as \ln
k_{t,\text{cutoff}}/Q$.
For the parameters above, this would give $n \sim 10^{8}$.
\logbook{}{Tried a run with a cutoff of
  $-207.5$ for $\as(Q) = 0.0016$ and got an average multiplicity of
  $150\text{k}$.
  \ttt{build-double/example-framework -no-obs -shower panlocal
    -alphas 0.0016 -lnvmin -491.25 -lnkt-cutoff -207.5 -nloops 2 -out
    a -use-diffs -output-interval 1 -spin-corr off -max-print 0}.
  Lowering the cutoff to $-230.5$ makes generation hard, but it looks
  like being $\order{640}\text{k}$ already.
  Extrapolating as $\ln N \sim |L|$
  with $L=-327.5$ would suggest something in the ballpark of $10^8$.
}
To resolve this issue, in addition to the fixed $k_t$ cutoff we also
use a dynamic cutoff. 
We track the shower scale $v_1$ of the first shower emission.
Knowing that we are restricting our attention to $\betaobs=\betaps$
observables and that standard global event shapes are recursively
infrared-and collinear safe~\cite{Banfi:2004yd}, we can be sure that
emissions with much lower values of $v$ will not modify the event
shapes.
Accordingly, when the shower reaches an $\ln v$ scale
that is sufficiently far below that of the first emission, we simply
stop showering.
That choice is specified by the \ttt{-dynamic-lncutoff -15} argument,
which stops the showering when $\ln v < \ln v_{1} + \delta_{\ln v} = \ln v_1 - 15$.
This ensures that the multiplicity is kept limited, e.g.\ for the
above run it averages to about $20$.

\begin{figure}[tp]
    \centering
    \mbox{
      \begin{subfigure}{0.48\textwidth}
        \includegraphics[page=1, width=\textwidth]{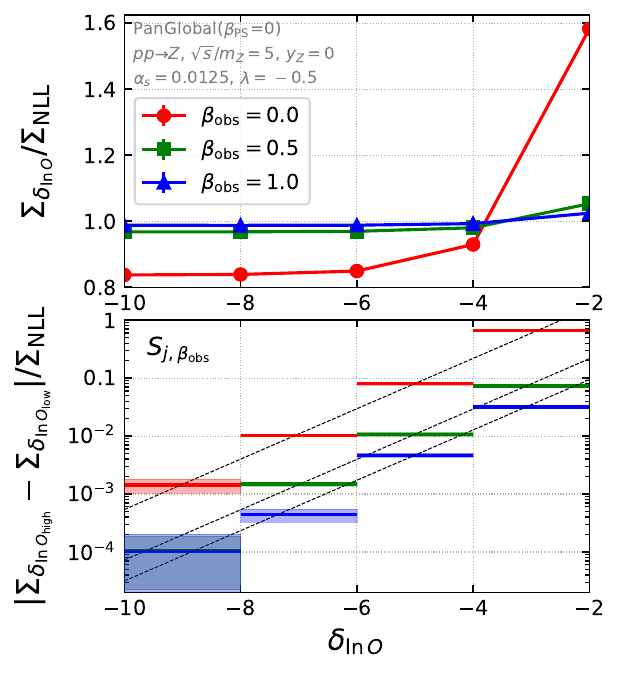}
        \caption{}
      \end{subfigure}
      \begin{subfigure}{0.48\textwidth}
        \includegraphics[page=2, width=\textwidth]{plots/plot-vetotest.pdf}
        \caption{}
      \end{subfigure}}
    \caption{Dependence of $\Sigma(\lambda)$ on the size, $\delta_{\ln
        O}$, of the parton-shower dynamic cutoff.
      The results are shown for the $S_{j, \betaobs}$ class of
      observables in Drell-Yan production, which sum $p_{ti}
      e^{-\betaobs|\eta_i|}$ over all final-state jets $i$ (with $p_t$
      defined as the transverse momentum with respect to the beam).
      The dependence on the size of the dynamic cutoff size is shown normalised to
      the NLL prediction for $\Sigma$, 
      for each of three
      values of $\betaobs$: 0 (red), 0.5
      (green) and 1.0 (blue).
      Two showers are shown
      (a) PanGlobal with $\betaps =
      0$ and (b) PanLocal (dipole) with $\betaps = 0.5$.
      The upper panels show the ratio of $\Sigma$ to the NLL result,
      showing the convergence as $\delta_{\ln O}$ is made more negative (note that the
      ratio is not expected to go to $1$, because of the finite value of $\as$).
      The lower panels show the difference between
      $\Sigma/\Sigma_\text{NLL}$ results for
      successive $\delta_{\ln O}$ values, again normalised to the NLL
      result, giving a clearer view of the degree of convergence.
      The dashed lines help illustrate that the behaviour is
      consistent with exponential dependence on $\delta_{\ln O}$.
    }
    \label{fig:veto-buffer-validation}
\end{figure}

To validate this approach, the dependence of the normalised cross section
$\Sigma(\lambda)$ on the size of the dynamic cutoff
is shown in
Fig.~\ref{fig:veto-buffer-validation}a (red line, corresponding to a $k_t$-like
observable, i.e.\ $\betaobs=0$).
It is illustrated for the $pp \to Z$ process, as carried out for the
studies of Ref.~\cite{vanBeekveld:2022ukn}, and one observes good
convergence of $\Sigma(\lambda)$ as the dynamic cutoff $\delta_{\ln
  v}$ is taken below $-9$ (for the $\betaps=\betaobs$ case discussed,
$\delta_{\ln v} \equiv \delta_{\ln O}$ in the plot).
The value of $-15$ that we use above leaves a comfortable margin.

The above procedure is sufficient when the angular power $\betaps$ in
the definition of the shower ordering variable coincides with the
angular power $\betaobs$ in the observable.
When $\betaps \neq \betaobs$ a more elaborate procedure is
needed.
As a starting point, we need to be able to evaluate the order of magnitude of the
contribution of each emission $i$ to the observable,
$O_{\text{approx},i}$.
Given knowledge of $\betaps$ and $\betaobs$, this can be determined
from the value of $\ln v$ and the (approximate) rapidity of the
emission.\footnote{With the help of the
  \ttt{ShowerBase::Element::lnobs\_approx(...)}  function in the
  \ttt{panscales} namespace.}
We then determine if the following condition holds
\begin{equation}
  \label{eq:veto-condition}
  \ln O_{\text{approx},i} < \max_{j<i}\{\ln O_{\text{approx},j}\} +
  \delta_{\ln O}
\end{equation}
where on the right-hand side the  max operation runs over prior accepted
emissions $j$ and $\delta_{\ln O}$ generalises the $\delta_{\ln v}$
discussed above.
If Eq.~(\ref{eq:veto-condition}) holds, then emission $i$ is discarded
and showering continues.
Showering is subsequently stopped when $\ln v$ is sufficiently small
such that for all emission rapidities one can be sure that
Eq.~(\ref{eq:veto-condition}) will always hold.

Note that for $\betaps \neq \betaobs$ this procedure is
guaranteed to be safe only for showers that respect the \panscales
condition that a given emission does not impact other emissions far in
the Lund plane.\footnote{%
  This condition is not satisfied for standard dipole showers.
  For $\betaobs \neq \betaps$ this results in super-leading logarithmic
  terms~\cite{Dasgupta:2020fwr} and such terms would not be fully
  reproduced with the above dynamic veto procedure.
}
Results are illustrated in Fig.~\ref{fig:veto-buffer-validation}, for
various combinations of $\betaps$ and $\betaobs$, showing
the relative change in $\Sigma$ between successive pairs of values of
$\delta_{\ln O}$, corresponding to the extremities of each horizontal
bar. 
One sees a behaviour that is consistent with an exponentially vanishing effect as
$\delta_{\ln O}$ becomes more negative.
Again, a choice of $-15$ should be more than adequate for NLL
logarithmic tests.
Note that the need for a more sophisticated dynamic veto and cutoff is
not the only challenge that arises with $\betaps \neq \betaobs$.
Further considerations are discussed below in
section~\ref{sec:weighted-unequal-beta}.

%......................................................................
\subsubsection{Weighted event generation for $\betaps = \betaobs$}
\label{sec:weighted-equal-beta}

Now we turn to the following part of the command line at the beginning
of section~\ref{sec:log-tests-global}
{\footnotesize%
\begin{verbatim}
    -weighted-generation -nev-cache 75000.0 
\end{verbatim}}
\noindent%
This is needed to address the fact that $\Sigma$ becomes
infinitesimally small in the limit $\as \to 0$ for fixed
$\lambda = \as \ln v$, specifically $\ln \Sigma \sim \lambda/\as$.
To get a more concrete sense of the challenge, consider that
$\Sigma \sim 10^{-103}$ for
$\lambda =-0.5$ and $\as=0.0016$ as in the command line at the start
of section~\ref{sec:log-tests-global}.
With unweighted event generation, it would take orders of magnitude
longer than the age of the universe to explore that region.

We address this challenge by greatly enhancing the number of events
whose first emission has an extremely small value of $\ln v$, assigning
a suitable weight to those events so as to reproduce the correct final
$\Sigma$ distribution.

We divide the full evolution range $\left[\ln v_{\max}, \ln v_{\min}\right]$
into a set of $n$ consecutive bins, each defined by their upper
boundaries, $\ln v^+_i$.
Writing the shower Sudakov form factor between two scales $v$ and $v'$
as $\Delta(v, v')$ (for the Born event), we precompute the part of the
Sudakov form factor associated with each bin.\footnote{This can be
  done either with a Monte Carlo integration (the default) or with
  Gaussian quadrature.
  The \ttt{-nev-cache} argument indicates the number of events used
  for the MC integration in each bin.
  We have found that a suitable choice is about $10\%$ of the total
  number of events that one wishes to generate.
}
The precomputed Sudakov can be inspected in the output file.

For each event, we choose a generation bin $i$ randomly, with a
probability $p_i$ that we take proportional to
$\ln v^+_i - \ln v^+_{i+1}$.
The shower  then starts from scale $v^+_i$.
If the first emission scale is above $v^+_{i+1}$, the shower continues down to
the dynamic cutoff as explained in section~\ref{sec:veto-buffer}.
If the first emission scale is below $v^+_{i+1}$, the emission is
discarded and showering is restarted from scale $v^+_i$, repeating
this until one generates an emission above $v^+_{i+1}$.
The weight assigned to the event is
\begin{equation} \label{eq:weighted-generation-weight}
    w = \frac{1}{p_i} \left(\Delta(v_{\max}, v^+_i) - \Delta(v_{\max}, v^+_{i+1}) \right).
\end{equation}
where the factor $1/p_i$ compensates for the likelihood of choosing
the window and the second factor accounts for the Sudakov form-factor
probability of having the first emission between $v^+_i$ and
$v^+_{i+1}$.

In practice we choose the bins such that $\ln v^+_i$ scales as
$-\sqrt{i}$, which in a fixed-coupling approximation ensures that the
Sudakov form factor decreases by a similar factor from one bin to the
next.
We take the total number of bins to be
$n = \ln \Delta(v^+, v^-)/\ln(u)$ where $u$ is a number of
$\mathcal{O}(1/2)$, so that the probability of each successive bin is
about half that of the previous bin.
This ensures that each attempt at starting the showering has a
$\sim 50\%$ chance of generating an emission within the given window,
while also ensuring that the event weight tracks the actual
first-emission Sudakov probability to within about a factor of two.

\begin{figure}[t]
    \centering
    \mbox{
        \begin{subfigure}{0.48\textwidth}
          \includegraphics[page=1, width=\textwidth]{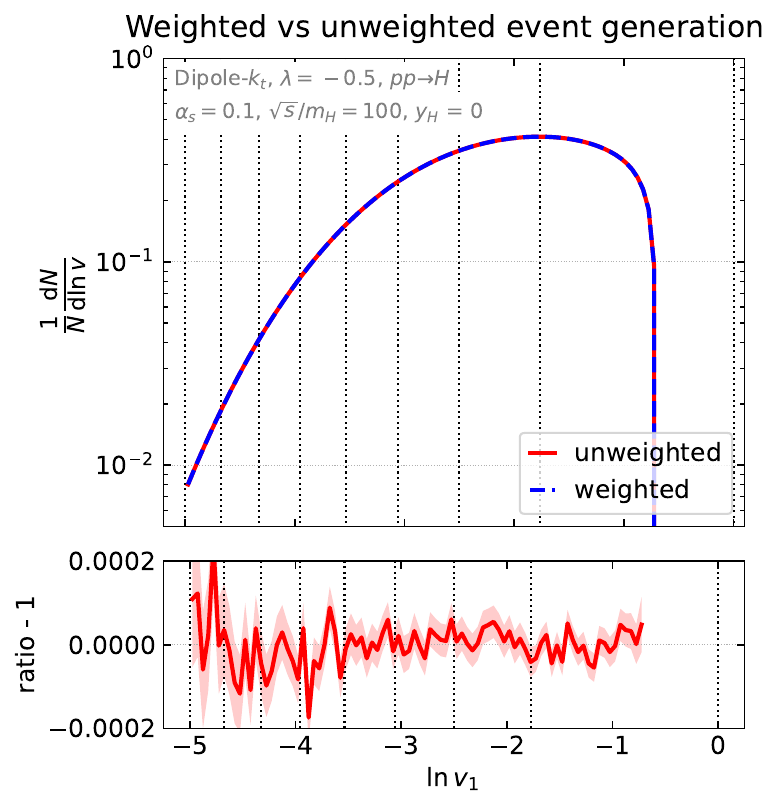}
          \caption{}
        \end{subfigure}
        \begin{subfigure}{0.48\textwidth}
          \includegraphics[page=2, width=\textwidth]{plots/plot-weighted-vs-unweighted.pdf}
          \caption{}
        \end{subfigure}
      }
    \caption{Validation of the weighted event generation.
      We show the results for Dipole-$k_t$ for $pp
      \to H$ collisions with $\sqrt{s}/m_H = 100$ and $y_H = 0$. We
      take $\as = 0.1$ and target $\lambda = -0.5$.
      (a) The distribution of $\ln v_1$ values
      for the first emission, with and without weighted generation,
      illustrating that they agree.
      The bottom panel shows
      the ratio between the weighted and unweighted results minus $1$,
      where the statistical uncertainty is indicated with the red
      band.
      (b) The relative statistical uncertainty in the $\ln v_1$
      distribution for both unweighted
      and weighted shower generation.
    }
    \label{fig:validation-weighted-vs-unweighted}
\end{figure}

Fig.~\ref{fig:validation-weighted-vs-unweighted}a shows a validation
of the correctness of the procedure for a physical value of the
coupling, $\as=0.1$, where it is straightforward to obtain high
accuracy with both weighted and unweighted approaches.
The plot shows the distribution of $\ln v_1$ for the first emission when
the generation is unweighted (red solid lines) and weighted (blue
dashed lines).
The two approaches agree to within statistical errors.
Fig.~\ref{fig:validation-weighted-vs-unweighted}b shows the size of
the relative statistical error in the two approaches, illustrating the
superiority of weighted generation for small $\ln v_1$ values, and
also illustrating that its statistical error is fairly independent of
$\ln v_1$, with a mild sawtooth structure that lines up with the generation
bin edges (indicated as vertical dotted lines).
% ......................................................................
\subsubsection{Weighted event generation for $\betaps \neq \betaobs$}
\label{sec:weighted-unequal-beta}
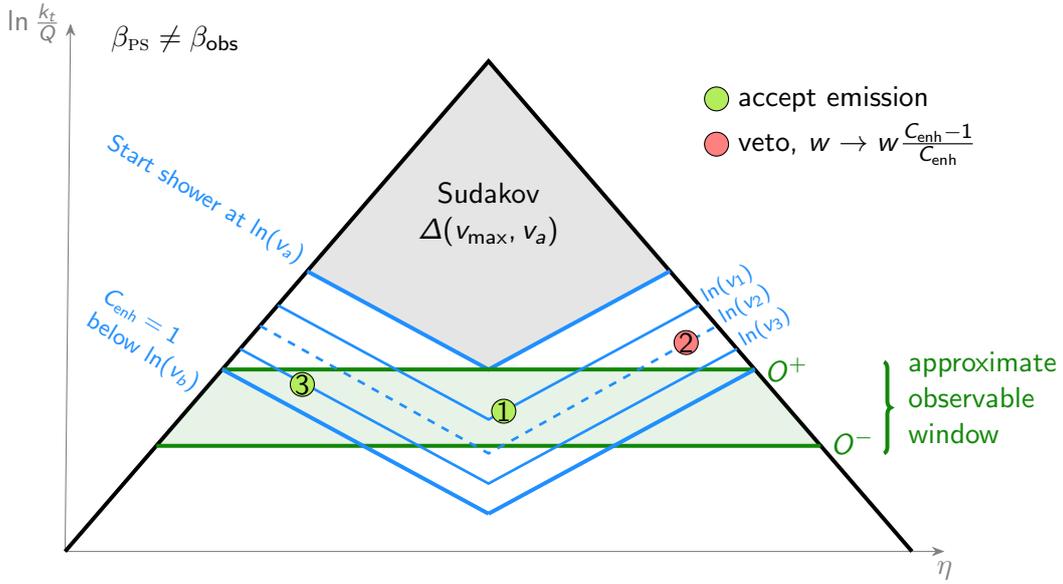
\begin{figure}\centering
  \pgfdeclarelayer{foreground}
  \pgfdeclarelayer{background}
  % tell TikZ how to stack them (back to front)
  \pgfsetlayers{background,main,foreground}
  % colours
  \definecolor{inchworm}{rgb}{0.7, 0.93, 0.36}
  \definecolor{indiagreen}{rgb}{0.07, 0.53, 0.03}
  \definecolor{dodgerblue}{rgb}{0.12, 0.56, 1.0}
  \newcommand{\colourobs}{indiagreen}
  \newcommand{\colourshower}{dodgerblue}
  \newcommand{\acceptcol}{inchworm}
  \newcommand{\vetocol}{red!50}
  \usetikzlibrary{arrows.meta}
  %\begin{tikzpicture}[decoration={brace}]
  \begin{tikzpicture}[decoration={brace},font=\sffamily\sansmath,every node/.style={scale=1.0}]
    %
    % define some coordinates
    \coordinate (bottom1) at (0.428571,0.5);
    \coordinate (top) at (6,7);
    \coordinate (bottom2) at (11.5714,0.5);
    % outer axis frame
    \coordinate [label={[color=gray,left]:$\ln\frac{k_t}{Q}$}](axtop) at (0.5,7.5);
    \coordinate [label={[color=gray,below]:$\eta$}](axbot) at (12,0.5);
    % draw the arrows
    \draw[-Stealth,color=gray] (bottom1) -- (axbot);
    \draw[-Stealth,color=gray] (bottom1) -- (axtop);
    % beta=betaobs label
    \coordinate[label={[right]:$\betaps\neq \betaobs$}] (axtopright) at ($(axtop)+(0.4,-0.2)$);
    % defines the lund triangle
    \draw[line width=1.5pt] (bottom1) -- (top) -- (bottom2) ;
    % define the sudakov
    \coordinate[label={[xshift=0.15cm,yshift=0.95cm,left,color=\colourshower]:\rotatebox{-31}{\footnotesize{Start shower at $\ln(v_a)$}}}] (valeft) at (3.61355,4.2158);
    \coordinate[label={[label distance=1.5cm, align=center]above:{Sudakov \\$\Delta(v_{\max},v_a)$}}](vamid) at (6, 2.9128);
    \coordinate[](varight) at (8.38645,4.2158);
    \draw[line width=1.5pt,color=\colourshower] (valeft) -- (vamid) -- (varight);
    %
    % fill in the upper Sudakov
    \begin{pgfonlayer}{background}
      \draw[fill opacity=0.2, fill=gray] (valeft) -- (vamid) -- (varight) -- (top) -- cycle;
    \end{pgfonlayer}
    % observable window, at kt/Q = 3.1 ends at kt/Q = 2
    \coordinate[label={[align=left,xshift=0.7cm,yshift=0.2cm,left,color=\colourshower]:\rotatebox{-31}{\parbox{2.4cm}{\footnotesize{$C_{\rm enh} = 1$ \\below $\ln(v_b)$}}}}] (obstopleft) at (2.49669,2.9128);
    \coordinate[label={[right, color=\colourobs]:{$O^+$}}] (obstopright) at (9.50331,2.9128);
    \coordinate[] (obsbotleft) at (1.62857,1.9);
    \coordinate[label={[right, color=\colourobs, yshift=0.05cm]:{$O^-$}}] (obsbotright) at (10.3714,1.9);
    \draw[fill=\colourobs, fill opacity=0.1] (obsbotleft) -- (obsbotright) -- (obstopright) -- (obstopleft) -- cycle;
    \draw[line width=1.5pt, color=\colourobs] (obsbotleft) -- (obsbotright);
    \draw[line width=1.5pt, color=\colourobs] (obstopleft) -- (obstopright);
    % accolade
    \draw [decorate,line width=1.5pt, color=\colourobs] ($(obstopright.north east)+(1.7,0.1)$) -- ($(obsbotright.south east)+(-10.3714+9.50331+1.7,-0.1)$);
    \coordinate [label={[align=left,color=\colourobs]:approximate \\ observable \\ window}] (obstext) at (12.5,1.8); 
    % place to stop the enhancement
    \coordinate[] (stopshowermid) at (6,1);
    \draw[line width=1.5pt, color=\colourshower] (stopshowermid) -- (obstopright);
    \draw[line width=1.5pt, color=\colourshower] (stopshowermid) -- (obstopleft);
    %
    % draw a few emsn lines
    %
    \coordinate[] (v2mid) at (6,2.25);
    \coordinate[label={[yshift=0.28cm,xshift=-0.15cm,right,color=\colourshower]:\rotatebox{31}{\scriptsize{$\ln(v_1)$}}}] (v2right) at (8.77345,3.7643);
    \coordinate[] (v2left) at (3.22655,3.7643);
    \draw[line width=1pt, color=\colourshower] (v2left) -- (v2mid) -- (v2right);
    \node [circle,fill=\acceptcol,inner sep=0pt,draw=black,minimum size=9pt] (v2ems) at (6.2,2.3592) {$1$};
    \coordinate[] (v3mid) at (6,1.8);
    \coordinate[label={[yshift=0.28cm,xshift=-0.15cm,right,color=\colourshower]:\rotatebox{31}{\scriptsize{$\ln(v_2)$}}}] (v3right) at (9.00701,3.49182);
    \coordinate[] (v3left) at (2.99299,3.49182);
    \draw[line width=1pt, dashed, color=\colourshower] (v3left) -- (v3mid) -- (v3right);
    \node [circle,fill=\vetocol,inner sep=0pt,draw=black,minimum size=9pt] (v3ems) at (8.6,3.2696) {$2$};
    \coordinate[] (v4mid) at (6,1.4);
    \coordinate[label={[yshift=0.28cm,xshift=-0.15cm,right,color=\colourshower]:\rotatebox{31}{\scriptsize{$\ln(v_3)$}}}] (v4right) at (9.26976,3.18528);
    \coordinate[] (v4left) at (2.73024,3.18528);
    \draw[line width=1pt, color=\colourshower] (v4left) -- (v4mid) -- (v4right);
    \node [circle,fill=\acceptcol,inner sep=0pt,draw=black,minimum size=9pt] (v2ems) at (3.55,2.714) {$3$}; %(3.5,2.765)
    %
    % legend
    \node [circle,fill=\acceptcol,inner sep=0pt,draw=black,minimum size=9pt] (aclab) at ($(axbot) + (-3,6)$) {};
    \coordinate [label={[right]:accept emission}] (acceptlabel) at ($(aclab)+(0.15,0)$);
    \node [circle,fill=\vetocol,inner sep=0pt,draw=black,minimum size=9pt] (vetlab) at ($(aclab)+(0,-0.6)$) {};
    \coordinate [label={[right]:veto, $w\to w \frac{C_{\rm enh}-1}{C_{\rm enh}}$}] (acceptlabel) at ($(vetlab)+(0.15,0)$);
  \end{tikzpicture}
  
  \caption{Illustration of some of the main steps in the weighted event-generation
    approach that is used for $\betaps \neq \betaobs$, for a specific
    target observable window.
    Beyond what is shown in the figure, one important element is that
    if, after showering, there are no emissions in the approximate 
    observable window, the event is discarded.
    The approach is additionally supplemented with the
    dynamic generation cutoff of section~\ref{sec:veto-buffer}. 
  }
  \label{fig:disequal-beta-weighting}
\end{figure}
The above weighted event-generation procedure is very powerful when
the Lund contour of the observable lines up with that of the shower
evolution variable, i.e.~$\beta_{\text{obs}} = \betaps$. If
this is not the case, a given value of the observable receives
contributions from different evolution windows with often widely
differing weights, significantly worsening statistical convergence.
In these situations, we employ a different procedure, which relates to
an approach that was first introduced in Ref.~\cite{Lonnblad:2012hz} in
the context of multi-jet merging.
The reader may wish to consult Fig.~\ref{fig:disequal-beta-weighting}
as they follow the description of the approach.
As with the dynamic cutoff in section~\ref{sec:veto-buffer}, it is
useful to introduce an approximate (soft-collinear) calculation of the
contribution to the observable from any given single emission,
$O_{\text{approx},i}$.
In a first instance the question is how to efficiently generate events
such that $O_{\max} \equiv\max_{i}\{O_{\text{approx},i}\}$ is in some
range $O^{-} < O_{\max} < O^{+}$, the green band in
Fig.~\ref{fig:disequal-beta-weighting}, labelled ``approximate
observable window''. 
Knowing the scaling of the observable, it is possible to analytically
work out the largest shower scale, $v_a$, that can generate an
approximate emission with observable-value $O^{+}$.
Showering then always starts from that scale $v_a$, with an initial weight
equal to $w = \Delta(v_{\max}, v_a)$, where $v_{\max}$ is the largest
kinematically accessible scale.
We then need to ensure that the showering does not generate any
emissions with $O_{\text{approx},i} > O^+$.
The simplest approach would be to veto every event that has any of the
emissions $i$ contributing such that $O_{\text{approx},i} >  O^+$.
This would in general lead to a very small fraction of surviving
events. 
Instead, and perhaps somewhat counter-intuitively, we use an
enhancement $C_\text{enh} > 1$ for the probability of generating
individual emissions.
If an emission $i$ has $O_{\text{approx},i} >  O^+$, the emission is discarded,
the event is kept, but the event weight $w$ is multiplied by a factor
such that
\begin{equation}
  \label{eq:w-Cenh-factor}
  w \to w \, \frac{C_{\text{enh}} - 1}{C_{\text{enh}}}.
\end{equation}
Emissions with $O_{\text{approx},i} \le  O^+$ (including those with
$O_{\text{approx},i} \le  O^-$) are instead accepted with
probability $1/C_{\text{enh}}$, with the event weight
unchanged.
Once $v$ goes below some value $v_b$ such that no further emission can
have $O_{\text{approx},i} > O^+$, the enhancement factor is set equal to $1$ and 
the shower 
continues down to the scale
of the dynamic cutoff.
After statistical averaging this gives exactly the same result as a
uniform-weight approach that discards every event with any emissions
$O_{\text{approx},i} > O^+$~\cite{Lonnblad:2012hz}.\footnote{One way of
   understanding this is to think that one starts with a number of
  replicas of the shower that will evolve in parallel.
  The number of replicas is equal to $C_\text{enh}$, and the
  enhancement of the emission probability can be interpreted as being
  equivalent to the total probability of emissions occurring in any of
  the replicas.
  Concentrating specifically on emissions with $O_{\text{approx},i} >  O^+$,
  the first time any of the replicas generates an emission with $O_{\text{approx},i} >
  O^+$, that replica is simply discarded, leaving
  $C_\text{enh}-1$ replicas. 
  The factor $ \frac{C_{\text{enh}} - 1}{C_{\text{enh}}}$ in
  Eq.~(\ref{eq:w-Cenh-factor}) is simply the ratio of surviving to
  original replicas.
  As the shower continues, the remaining weight
  $ \frac{C_{\text{enh}} - 1}{C_{\text{enh}}}$ is then shared back out
  across $C_\text{enh}$ replicas again, and so the procedure
  continues.  }
Finally, events are only accepted if at least one emission has an
approximate observable value
in the range $O^{-} < O_{\max} < O^{+}$.

The width of the event weight distribution gets smaller as $C_{\text{enh}}$
is made larger, but there is an associated slow-down of the showering,
because of the increased emission probability.
The optimal choice for $C_{\text{enh}}$ involves some balance between
these two aspects.
In practice we use $C_{\text{enh}} = 20$ for showers with
$\betaps = 0$, and $C_{\text{enh}} = 100$ for showers with
$\betaps = 0.5$.

The final step is that for a given value of the constraint on the
actual observable (e.g.\ $\as \ln O < \lambda = -0.5$),
we need to add together several contiguous $O^{\pm}$ windows
above the constraint, plus one final window without a lower bound
(i.e.\ $O^-=0$).
The need for several windows is because the actual value of the
observable can be smaller than the approximate value of the
observable by some $\order{1}$ factor.

In its current form, the code
(\mainlink{analyses/nll-validation/shower-global-obs.cc}) performs
separate runs for each of several windows and the python script that
manages the calculations (\analysislink{run-nll-tests.py}) adds the
results together.
A command line that sets up the use of the above algorithm is, for instance

{\footnotesize%
\begin{verbatim}
  build-double/shower-global-obs -Q 1.0 -no-spin-corr  -nloops 2 -alphas 0.008 \
        -shower panglobal -beta 0.0 -colour NODS \
        -lambda-obs -0.5 -beta-obs 0.5 -lnkt-cutoff -77.5 \
        -strategy RouletteEnhanced -enhancement-factor 20.0 \
        -ln-obs-buffer 3.5 -nln-obs-div 7 -veto-buffer -15.0 \
        -use-diffs  -rseq 11 -nev 70000 \
        -out nll-tests-tmp/panglobal00-fapx0.5-lambda0.5-as0.008-NODS-rseq11.dat
\end{verbatim}

}
\noindent The second and third lines set up a shower and an observable class
with different $\betaps$ and $\betaobs$ values.
The fourth line sets up the strategy described above and the
associated value of the enhancement factor, corresponding to
$C_{\text{enh}}$.
The fifth line indicates that $7$ approximate observable windows are
explored, extending to $e^{3.5}$ times above the target value of
the observable.
The code performs runs in each of the $7$ separate
approximate-observable windows, producing one output file for each
(Fig.~\ref{fig:disequal-beta-weighting} corresponds to a single
window).
The \ttt{-veto-buffer -15} argument on the fifth line sets the size
of the dynamic veto and cutoff, as explained for $\betaps \neq \betaobs$ at the
end of section~\ref{sec:veto-buffer}.
If a user wishes to explore new observables or new showers, they are
strongly advised to manually verify that the contribution to the cross
section from the highest window is sufficiently suppressed relative to
the total result.
This can be done by examining the output files from the above command,
there being one file per observable window.

% ----------------------------------------------------------------------
\subsection{Single-logarithmic observables, e.g.\ non-global logarithms}
\label{sec:non-global-log-tests}

An example of a single-logarithmic test, such as the transverse
energy flow in a rapidity slice in $e^+ e^- \to q\bar{q}$ collisions,
can be performed with the
\showercodelink{example-nonglobal-nll-ee.py} script in the
\mainlink{shower-code/} directory.
This will execute the following command

{\footnotesize%
\begin{verbatim}
  build-doubleexp/example-nonglobal-nll-ee -Q 1.0 -shower panglobal -beta 0.0 \
        -nloops 2 -colour CATwoCF  \
        -slice-maxrap 1.0 -lambda-obs-min -0.5 \
        -alphas 1e-09 -lnkt-cutoff -501000000.0 -ln-obs-margin -11 \
        -strategy CentralRap -half-central-rap-window 10  \
        -spin-corr off -nev 10000 -rseq 11 \
        -out example-results/lambda-0.5-alphas1e-09-rseq11.res
\end{verbatim}}
\noindent The second line indicates that a 2-loop running coupling
is to be used and that the colour scheme is a large-$\nc$ scheme in
which $C_A = 2 C_F = 8/3$.\footnote{Recall that the PanScales
  showers are logarithmically accurate for non-global logarithms only in
  the large-$\nc$ limit, though the subleading-colour
  schemes~\cite{Hamilton:2020rcu} are numerically close to the
  full-colour results\cite{Hatta:2013iba,Hatta:2020wre}.
  The schemes of Ref.~\cite{Hamilton:2020rcu} can be obtained by
  replacing \ttt{CATwoCF} with \ttt{NODS} or \ttt{Segment}.
}
The third line indicates the size of the rapidity slice in which
the in-slice energy flow $E_t$ will be measured, and sets the minimum
value of $\lambda = \as \ln E_t/Q = -0.5$.

The fourth line indicates that we run at an infinitesimal value of the
coupling, $\as(Q) = 10^{-9}$ (still working in units of
$\sqrt{s}\equiv Q = 1$), with a sufficiently small $k_t$ cutoff,
$e^{-501000000}$, so as to cover the region down to $\lambda = -0.5$.
The point of using such a small value of $\as$ is to avoid a
substantial extrapolation to $\as = 0$, alleviating the need for runs
at multiple values of $\as$.
It is physically possible to use such an infinitesimal value because
the observable $\Sigma(\lambda,\as)$ is independent of $\as$ for $\as \to
0$.
This is in contrast with the case of global event shapes where
$\ln \Sigma(\lambda,\as) \sim {\lambda/\as}$.
The huge range of orders of magnitude of momenta requires the use of the
\texttt{doubleexp} type.

As with the case of global event-shape tests, a straightforward run
with the above parameters would give multiplicities that are much too
high to be managed (in fact, the situation is even worse, because of
the much smaller value of $\as$).
The mitigation procedure is different in this case: the combination of
arguments on the fifth line causes the shower to only generate
emissions within a window where the absolute rapidity is less than $11$
(with respect to the emitter or spectator).
For showers that satisfy the \panscales conditions, as long as
$\lambda$ is not too large and the window is large enough, the results
should be (and are) independent of the window size.
Corresponding arguments also exist for an analogous modification of the
shower to generate only hard-collinear emissions, as is relevant for
many spin-correlation, fragmentation function and PDF evolution tests. 

Note that for verifying next-to-single-logarithmic
accuracy~\cite{Banfi:2021owj,Banfi:2021xzn,Becher:2023vrh} for
non-global logarithms~\cite{FerrarioRavasio:2023kyg} it is necessary
to enable the double-soft matrix element corrections and associated
virtual corrections (with the \texttt{-double-soft} command-line
argument, available only for the PanGlobal showers with $\beta=0$ and
$\beta=0.5$, for $e^+e^-$ collisions).
One also has to modify the above command line to run at multiple small
but finite values of $\as$, so as to be able to determine the first
derivative of $\Sigma(\lambda, \as)$ with respect to $\as$ in the
$\as \to 0$ limit.
Doing so accurately requires considerable computer resources.
Users who wish to explore this are advised, in a first instance, to
run with the \texttt{-split-dipole-frame} option for the PanGlobal
showers with $\betaps=0$, which is the most computationally efficient
setup.

%======================================================================
\section{Implementing a new shower}
\label{sec:new-shower}

The \panscales framework allows for relatively straightforward
addition of new dipole and antenna
showers with alternative kinematic maps or ordering
variables.
This allows the user to leverage the existing code for colour and spin
handling, as well as the infrastructure for logarithmic accuracy tests
and the interface to \pythia.
As an example, our own toy implementation of the \pythia final-state shower
involves about $250$ lines of header (\showercodelink{ShowerToyPythia8.hh}),
most of which are boilerplate code, and a further $250$ lines in
\showercodelink{ShowerToyPythia8.cc}, more than half of which are comments or
blank lines.
For a slightly more elaborate example that handles also initial-state
radiation, the user may wish to look at the \ttt{ShowerDipoleKt}
class.

Here we outline some of the aspects that a user should keep in mind in
implementing their own new shower.
Firstly, the code is in the \ttt{panscales} namespace.
Inspecting the code, the user will see that rather than \ttt{double},
many variables are in \ttt{precision\_type}: this corresponds to the
precision that was chosen in the build step.
Generically, \ttt{double} is used for logarithmic variables and
acceptance probabilities, while \ttt{precision\_type} is to be used
for (non-logarithmic) kinematic variables and associated
matrix-element calculations, where rounding errors and large exponents
may be encountered.

\begin{figure}[h!t!]
  \centering
\tikzstyle{arrow} = [thick,->,>=stealth, line width=0.5mm]
\tikzstyle{startstop} = [rectangle, rounded corners, minimum width=4cm, minimum height=0.5cm,text centered, draw=black, text width=5cm]
\usetikzlibrary{backgrounds}
% define layers
\pgfdeclarelayer{foreground}
\pgfdeclarelayer{background}
% tell TikZ how to stack them (back to front)
\pgfsetlayers{background,main,foreground}
%\begin{tikzpicture}[node distance=1.3cm]
\begin{tikzpicture}[node distance=1.3cm,font=\sffamily\sansmath,every node/.style={scale=0.9}]
\node (start) [startstop] {
  Choose \ttt{lnv} and element
};
\node (start-user) [startstop , right of=start, xshift=6cm, yshift=-0.5cm] 
       {Return \ttt{lnb} range};
\node (lnb) [startstop, below of=start] {
  Choose \ttt{lnb}
};
\node (lnkt) [startstop, below of=start-user] {
  Return \ttt{lnkt} for $\alpha_s$
};
\node (as) [startstop, below of=lnb] {
  $\alpha_s^{\rm CMW}(k_t^2)$ acceptance
};
\node (accept-user) [startstop, below of=lnkt] 
      {Determine branching\\weights};      
\node (accept) [startstop, below of=as] 
      {Decide whether to branch, select channel};
\node (phi) [startstop, below of=accept,yshift=-0.2cm] 
      {Generation \ttt{phi} and spin acceptance};
\node (abovephiarrow) [above of=phi]{};
\node (belowphiarrow) [below of=phi]{};
\node (dokin-user) [startstop, right of=phi, xshift=6cm] 
      {Generate dipole kinematics};
\node (colour) [startstop, below of=phi,yshift=-0.02cm] 
      {Apply colour corrections};
\node (update-base) [startstop, below of=dokin-user,yshift=-1cm] 
     {Standard event update (\ttt{ShowerBase::Element})};
\node (update-event) [startstop, below of=update-base,yshift=-0.2cm] 
     {Shower-specific event updates (e.g.\ boosts)};
\node (finalise) [startstop, below of=colour, yshift=-2cm] 
     {Final bookkeeping};
     
%%%%%%%%%%%%% big boxes
\node (showerrunner) [rectangle, rounded corners, minimum width=5.7cm, minimum height=11cm,  draw=black, below of=accept, yshift=0.5cm, text centered,
 text depth = 11cm] 
      {\ttt{ShowerRunner}};
\node (user) [rectangle, rounded corners, minimum width=5.7cm, minimum height=11cm, yshift=-0.3cm, draw=black, below of=accept-user,text centered,
text depth = 11cm] 
    {\ttt{ShowerUserDefined::Element}};
% arrows

\begin{pgfonlayer}{background}
\draw [arrow] ($(start.east)+(0,-0.1)$) -- ($(start-user.west)+(0,0.1)$);
\draw [arrow] ($(start-user.west)+(0,-0.1)$) -- ($(lnb.east)+(0,0.1)$);
\draw [arrow] ($(lnb.east)+(0,-0.1)$) -- ($(lnkt.west)+(0,0.1)$);
\draw [arrow] ($(lnkt.west)+(0,-0.1)$) -- ($(as.east)+(0,0.1)$);

\draw [arrow] ($(as.east)+(0,-0.1)$) -- ($(accept-user.west)+(0,0.1)$);
\draw [arrow] ($(accept-user.west)+(0,-0.1)$) -- ($(accept.east)+(0,0.1)$);

\draw [arrow] (accept) -- (phi);
\draw [arrow] ($(dokin-user.west)+(0,-0.1)$) --  ($(phi.east)+(0,-0.1)$);
\draw [arrow] ($(phi.east)+(0,0.1)$) --  ($(dokin-user.west)+(0,0.1)$);
\draw [arrow] (phi) --  (colour);

\draw [arrow] ($(colour.east)+(0,-0.1)$) -- ($(update-base.west)+(0,0.1)$);

\draw [arrow] ($(update-event.west)+(0,-0.1)$) -- ($(finalise.east)+(0,0.1)$);
\draw [arrow] (update-base) --  (update-event);
\end{pgfonlayer}
\begin{pgfonlayer}{foreground}
\node (emissioninfo) [rectangle, rounded corners, minimum width=0.8cm, minimum height=10.6cm,  yshift=0.5cm, draw=black,  fill=gray!50, right of=accept, below of=accept, text centered, xshift=2.35cm] 
{\rotatebox{90}{\ttt{ShowerBase::EmissionInfo}}};
\end{pgfonlayer}
\end{tikzpicture}
\caption{Simplified structure of a single trial for an emission in the
  shower.
  Steps on the left are performed by the centralised \ttt{ShowerRunner} class.
  Shower-specific steps, shown on the right, are the responsibility of the
  \ttt{ShowerUserDefined} implementation, as well as its sub-classes.
  The  \ttt{EmissionInfo} sub-class stores all required information needed
  for the current branching, such as the splitting variables and
  constructed kinematics.
  The emitting dipole is accessible in the \ttt{Element} sub-class, which contains
  the majority of the shower-specific code.
  The matching, implementation of the spin correlations throughout the event
  evolution, and the double-soft splitting functionalities are not
  illustrated here.
}
\label{fig:showerrunner}
\end{figure}
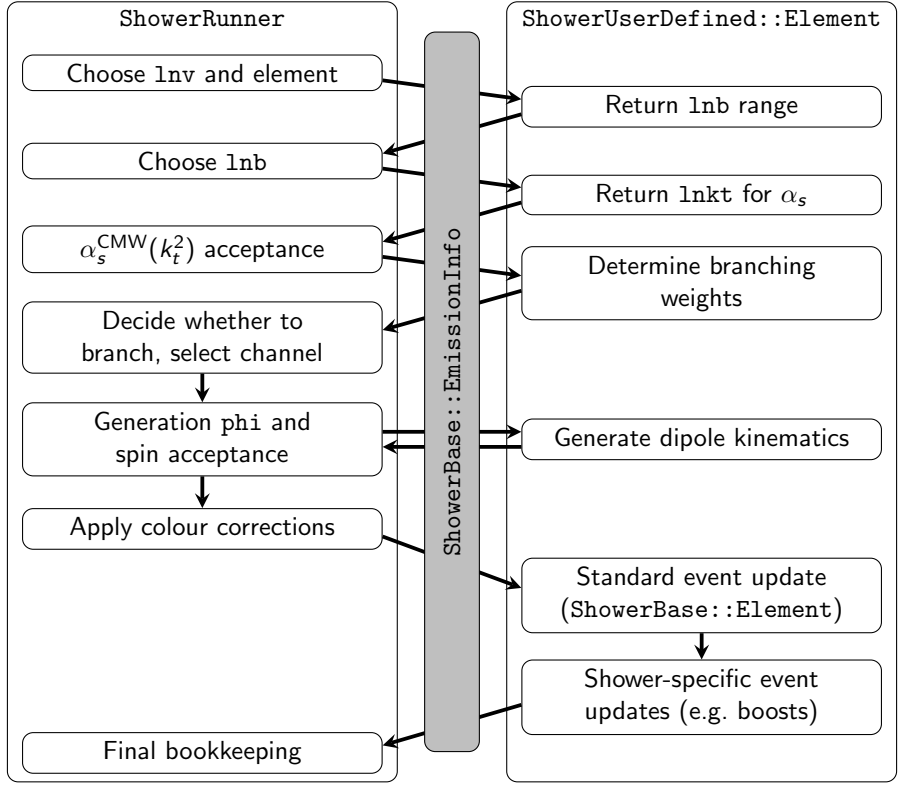

Much of the core work of showering is carried out by the
\ttt{ShowerRunner} class, which is essentially agnostic as to the specific
details of any given shower. 
The basic work flow of \ttt{ShowerRunner} is depicted in Fig.~\ref{fig:showerrunner}.
The \ttt{ShowerRunner} class is constructed by passing a pointer to any
class that derives from \ttt{ShowerBase}, whose role is to handle a
small, well-defined set of shower-specific tasks, such as providing
the acceptance probability and implementing the shower's kinematic
map. 
The functions that the user should provide to the shower
are also shown in Fig.~\ref{fig:showerrunner}.
For concreteness, with this distribution, we have supplied two files,
\showercodelink{ShowerUserDefined.hh} and \showercodelink{ShowerUserDefined.cc}, derived
from \ttt{ShowerBase}, which one can modify.
The functions that need to be 
touched are marked with ``\ttt{USER-TODO}'' in the code.
The user-defined shower can then be run with the \ttt{-shower
  user-defined} command line argument.
Note that the template is meant for $e^+ e^-$ showers.
For additional features like initial-state radiation, matching or the
implementation of double-soft currents, the reader should examine the
structure of other existing showers.

The class \ttt{ShowerUserDefined} implements:
\begin{itemize}
\item Member functions that provide textual descriptive information about the
  shower.
\item Member functions that provide structural information about the shower,
  notably whether the shower is a dipole shower (only the emitter
  splits), or an antenna shower (both the emitter and spectator can
  split) via the function \ttt{only\_emitter\_splits()}, and
  \ttt{n\_elements\_per\_dipoles()} (two for a dipole shower, one for
  an antenna).
  For showers with a global kinematic map, one where a dipole branching
  also impacts the kinematics of particles not in the dipole, the function
  \ttt{is\_global()} should return \ttt{true} (more on this later).
\item Member functions to create the two sub-classes
  \ttt{ShowerUserDefined::EmissionInfo}, and  \ttt{ShowerUserDefined::Element}.
\end{itemize}
The two sub-classes encode the data associated with an emission and
most of the implementation of the shower splitting:
\begin{itemize}
\item \ttt{ShowerUserDefined::EmissionInfo} derives from \ttt{ShowerBase::EmissionInfo}.
  A pointer to its base class is used by \ttt{ShowerRunner} as the main structure to keep track
  of the emission through the various steps of constructing that
  emission, and is generally passed to all shower-specific functions.
  Much of the required functionality is already present in the base
  class.
  Users may extend it to cache quantities computed
  at specific steps of a single emission that is trialled, so as to not
  reproduce them again at a later stage.
   An example would be the absolute $k_t$ of an emission, or the
   longitudinal momentum fraction $z$,
   which might first be calculated in the early stages of
   Fig.~\ref{fig:showerrunner} and then used later when generating the
   new dipole kinematics.
  
\item \ttt{ShowerUserDefined::Element} derives from
  \ttt{ShowerBase::Element}.
  It is responsible for carrying out almost all of the
  shower-specific work.
  In case of multiple types of elements (i.e.~II, IF and FF dipoles),
  one might choose to derive distinct element classes for each,
  typically each from an intermediate \ttt{ShowerUserDefined::Element} base
  class (see e.g.~\ttt{ShowerDipoleKt}).
\end{itemize}
Let us focus here on the \ttt{Element} class.
For a dipole shower, where each dipole end is associated with a
distinct kinematic map, there will be two \ttt{Element}s per dipole,
one for each end.
For an antenna shower, there will be just one \ttt{Element} per dipole.
The branching kinematics are governed by three variables:
\begin{itemize}
\item $\ln v$, the logarithm of the (dimensionful) ordering variable.
\item $\ln b$, a logarithmic variable for the longitudinal degree of
  freedom of the emission. In the soft-collinear region of a
  back-to-back dipole it might, for example, map directly to a
  rapidity, or to $\ln z$ where $z$ is the radiated particle's
  momentum fraction.
\item $\phi$, the azimuthal angle for the emission.
\end{itemize}
The main non-trivial member functions that need to be implemented are
the following:
\begin{itemize}
\item \ttt{lnb\_extent\_const()} and \ttt{lnb\_extent\_lnv\_coeff()}:
  for any given $\ln v$, \ttt{ShowerRunner} will take the $\ln b$
  generation range to be taken equal to
  \begin{equation}
    \label{eq:kin-range}
    \text{\ttt{lnb\_extent\_const()}} + \ln v \times \text{\ttt{lnb\_extent\_lnv\_coeff()}}
  \end{equation}
  where typically the first term would be a function of the dipole
  kinematics (as encoded in the corresponding \ttt{Element}) and the
  second term would not.
  The use of a simple analytic form for the extent facilitates the
  generation of the $\ln v$ distribution.
  
\item \ttt{lnb\_generation\_range(lnv)}:  for a given value of the
  evolution scale \ttt{lnv}, returns a \ttt{Range} object containing
  the limits of $\ln b$ generation.
  The difference between the upper and lower limits should coincide
  with the extent expected from the previous bullet point.
  Not all $\ln b$ values in the range need be kinematically valid.

\item \ttt{lnv\_lnb\_max\_density()}: returns the maximal emission
  density (for a $d\!\ln v \, d\!\ln b \,\frac{d\phi}{2\pi}$ measure).
  Typically this would be the soft-collinear limit of the emission
  density
  \begin{equation}
    \frac{dP}{d\!\ln v\, d\!\ln b \,d\phi/2\pi} = \frac{C_A \as^{\max}}{\pi}
  \end{equation}
  where $\as^{\max}$ is the maximal value that the strong coupling can
  take.\footnote{%
    All current shower implementations have a private member pointer
    \ttt{Element::\_shower}, and the base shower class has a
    \ttt{max\_alphas()} member, as well as a \ttt{qcd()} member that
    supply access respectively to $\as^{\max}$ and QCD constants.
    Together these provide the information needed to calculate the
    maximum density.
  }

\item \ttt{lnkt\_approx(lnv,lnb)}: returns the logarithm of the
  transverse momentum of an emitted parton with that $\ln v$, $\ln b$
  combination.
  The result should be exact in the soft-collinear limit, but
  does not need to be exact in the soft large-angle or in the
  hard-collinear limits.
  It is typically used for the evaluation of $\alpha_s(k_t)$,
  as well as for some of the dynamic emission vetoing used for
  logarithmic accuracy tests.
  
\item \ttt{eta\_approx(lnv,lnb)}: similar, but returns the
  rapidity of the emitted parton in the soft-collinear limit,
  used in the computation of colour transition points and dynamic
  emission vetoing.
  
\item \ttt{acceptance\_probability(emission\_info)}: sets information
  subsequently used by \ttt{ShowerRunner} in order to calculate
  the probability that the dipole splits.
  It makes use of the generation variables \ttt{lnv} and \ttt{lnb}, as
  cached in \ttt{emission\_info}.
  Specifically it sets the following member variables in \ttt{emission\_info}:
  \begin{itemize}
  \item \ttt{emitter\_weight\_rad\_gluon}   : weight for the emitter to radiate a gluon;
  \item \ttt{emitter\_weight\_rad\_quark}   : weight for the emitter to radiate a (anti-)quark;
  \item \ttt{spectator\_weight\_rad\_gluon} : weight for the spectator to radiate a gluon;
  \item \ttt{spectator\_weight\_rad\_quark} : weight for the spectator to radiate a \mbox{(anti-)quark.}
  \end{itemize}
  The weights to radiate a gluon/quark depend on the
  splitting functions.
  Only for an antenna shower will the spectator weights be non-zero.
  They are to include a factor of the maximal allowed value of the
  strong coupling $\as^{\max}$,
  with the \ttt{ShowerRunner} class then accounting for an
  $\as(k_t)/\as^{\max}$ factor.
  In initial-state branchings, \ttt{ShowerRunner} also accounts for an
  appropriate PDF ratio factor.
  The user may choose to use the splitting functions implemented in
  \showercodelink{QCD.hh}. This can be done by calling
  \ttt{fill\_dglap\_splitting\_weights}, which needs the longitudinal
  radiated momentum fraction $z$ (where $z\to 0$ indicates the soft
  limit).
  In addition, when using spin correlations, the user
  should set
  \begin{itemize}
  \item \ttt{z\_radiation\_wrt\_emitter} : collinear momentum fraction with respect to the emitter;
  \item \ttt{z\_radiation\_wrt\_spectator}: collinear momentum fraction with respect to the spectator.
  \end{itemize}
  The \ttt{acceptance\_probability(...)} function returns a
  \ttt{bool}, where \ttt{false} indicates that the generation
  variables are definitely outside the kinematic limit.
  For most showers, if it returns \ttt{true} then the generation
  variables would normally be inside the kinematic limit.
  The emitter and spectator splitting probabilities are then used later
  in \ttt{ShowerRunner} to accept/reject a splitting, and choose
  the splitting channel.

\item \ttt{do\_kinematics(emission\_info, rp)}: constructs the post-branching momenta of the
  emitter, the spectator and the newly radiated particle. 
  For this, one may again use the cached generation variables \ttt{lnv}, \ttt{lnb} and
  \ttt{phi}, alongside any other quantity that the user stored in \ttt{emission\_info}.
  The post-branching momenta should 
  be stored in \ttt{emission\_info} under the names
  \ttt{emitter\_out}, \ttt{spectator\_out} and \ttt{radiation}. 
  This function returns a \ttt{bool}, where \ttt{true} indicates that
  the generation variables were inside the physical kinematic
  limit.\footnote{%
    It is possible for the \ttt{acceptance\_probability(...)}
    function to return true even outside the kinematic limits.
    In this case there are two possible avenues for imposing the
    kinematic limit.
    One is for the shower's \ttt{Element} class to overload the
    base-class member function
    \ttt{check\_after\_acceptance\_probability(...)}, which gets
    called after $\as$ and PDF factors have been incorporated into the
    branching probabilities and those have been used to decide to
    continue with the emission generation.
    It is called before $\phi$ and the splitting channel are known.
    Alternatively \ttt{do\_kinematics(...)} is called with knowledge
    of the $\phi$ value and channel, and can return \ttt{false} if the
    emission is outside the kinematics.
    The former is the only route that is currently valid in order for
    spin correlations to be correctly accounted for.
  }
  Note that the pre-branching emitter and spectator momenta should be
  taken from a variable \ttt{rp} of type \ttt{RotatedPieces}.
  This class is part of the framework for handling directional
  differences.
  It provides a rotated version of the dipole with one or other of its
  particles aligned along the $z$ axis, which allows the user to
  retain high precision in the branching kinematics (specifically,
  small components along the $x$ or $y$ axis) without explicit
  knowledge of the underlying direction-difference structures.
  The direction-difference infrastructure then takes care of deducing
  the correct momenta and direction differences in the original frame.

\item \ttt{update\_event(...)}:  
  In the \ttt{Element} base class, the \ttt{update\_event(...)}
  member function takes care of replacing the
  pre-branching emitter and spectator particles with the post-branching ones,
  adds the radiated particle to the event, and takes care of some of
  the bookkeeping associated with colour handling.
  However, other particles in the event are by default not modified. 
  Therefore, for showers with a global kinematic map, this function
  needs to implement additional operations on the rest of the event
  (e.g.\ a boost or rescaling).
  These would typically be preceded by an explicit call to
  \ttt{ShowerBase::Element::update\_event(...)}.
  The event is then further processed by
  \ttt{ShowerRunner}, updating the event dipoles, colour and
  spin-density structure in addition to any caching associated with the
  event generation.
  Note that cached information associated with dipoles other than the
  newly-created dipole and the splitting dipole are by default not
  updated, unless the \ttt{ShowerUserDefined::is\_global()} function
  returns true.
\end{itemize}
Once implemented, the new shower can be run by using the flag
\ttt{-shower user-defined}.

If the user would like their shower to work with direction
differences, they should inspect how this is implemented in existing
showers.
Aside from the \ttt{RotatedPieces} discussion above, calculations of
dot products in determining dipole invariants should make explicit use
of knowledge of direction differences (available from the
\ttt{dirdiff\_3\_minus\_3bar} member variable of
\ttt{element.dipole()}), and \ttt{do\_kinematics(...)}  should use
$3$-vectors in its internal calculations to avoid triggering
off-mass-shell errors.
Furthermore if the shower carries out any global boosts, these need to
be performed in a way that correctly boosts also the full set of
dipole direction differences.
The user is invited to inspect the code of existing showers for
further details.

It is important also to test the correctness of the
direction-differences implementation.
Typically we do this by first running a double-precision build with a
physically sensible range and verifying that results are identical
with/without the \ttt{-use-diffs} argument.
Then we create a build in the \ttt{doubleexp} type, running with
\ttt{-use-diffs} and a logarithmic range of about $1000$ (and
correspondingly low $\as$, so that the multiplicity stays of the order of
$10{-}100$) and compare the output to a build with the much slower
\ttt{mpfr4096} type, running without \ttt{-use-diffs}.
Again the results should be identical, though typically a few
iterations are likely to be necessary to identify all sources of potential
loss of precision.
A final comment is that, by default, the logarithmic-accuracy tests of
section~\ref{sec:running-log-tests} run in double precision, with a
$\ln v$ range reaching about $300$.
However, for this to work, the shower should not ever do more than
take the square of a momentum, otherwise the result will exceed the exponent
that can be represented in double precision.
If this is a problem, the user should modify a configuration flag in
\showercodelink{example-global-nll-ee.py} so as to use \ttt{doubleexp} (which is
somewhat slower).

%======================================================================
\section{Conclusions}
\label{sec:outlook}

This 0.1 series release of the \panscales{} code allows users to start exploring
its features and techniques, notably for tests of logarithmic accuracy
of parton showers.
It also demonstrates an early version of the interface to
\pythia.
While we do not yet recommend its use for phenomenological
applications, we hope that this early release of the code will provide
a foundation for exploring connections with other projects, so as to
enable the wide ecosystem of collider physics tools to benefit from
the validated logarithmic accuracy of the \panscales showers.

\section*{Acknowledgements}

We thank Fr\'ed\'eric Dreyer for his contributions to the \panscales
project and code during the initial stages of the project.
We are grateful to Peter Skands and Silvia Zanoli for testing a
pre-release version of the code and for helpful comments.

This work has been funded by the European Research Council (ERC) under
the European Union's Horizon 2020 research and innovation program
(grant agreement No 788223),
by a Royal Society Research Professorship
(RP$\backslash$R1$\backslash$180112, GPS and LS, and
RP$\backslash$R$\backslash$231001, GPS) and by the Science and
Technology Facilities Council (STFC) under grants ST/T000864/1 (MvB,
GPS), ST/X000761/1 (GPS), ST/T000856/1 (KH) and ST/X000516/1 (KH),
ST/T001038/1 (MD) and ST/00077X/1 (MD).
LS is supported by the Australian Research Council through a Discovery
Early Career Researcher Award (project number DE230100867).
The work of PM is funded by the European Union (ERC, grant agreement
No. 101044599). Views and opinions expressed are however those of the
authors only and do not necessarily reflect those of the European
Union or the European Research Council Executive Agency. Neither the
European Union nor the granting authority can be held responsible for
them.
We also thank each others' institutes for hospitality during the
course of this work.

\bibliography{MC.bib}

\nolinenumbers

\end{document}